\documentclass[aps,prl,twocolumn,groupedaddress,showpacs]{revtex4-1}

\usepackage{amsmath}
\usepackage{SIunits}
\usepackage{graphicx}
\usepackage{epstopdf}
\begin{document}

\title{Towards a Mg lattice clock: Observation of the $^1S_{0}-$$^3P_{0}$ transition and determination of the magic wavelength}
\author{A. P. Kulosa$^1$, D. Fim$^1$, K. H. Zipfel$^1$, S. R\"{u}hmann$^1$, S. Sauer$^1$, N. Jha$^1$, K. Gibble$^{1,2}$, W. Ertmer$^1$, and E. M. Rasel$^1$}
%\thanks{}
\affiliation{$^1$Institut f\"{u}r Quantenoptik, Leibniz Universit\"{a}t Hannover, Welfengarten 1, 30167 Hannover, Germany\\
	$^2$Department of Physics, The Pennsylvania State University, University Park, Pennsylvania 16802, USA}
\author{M. S. Safronova$^{1,2}$, U. I. Safronova$^3$, S. G. Porsev$^{1,4}$}
\affiliation{$^1$Department of Physics and Astronomy, University of Delaware, Newark, Delaware 19716, USA\\
	$^2$Joint Quantum Institute, NIST and the University of Maryland, College Park, Maryland 20899, USA\\
	$^3$Department of Physics, University of Nevada, Reno, Nevada, 89557, USA\\
	$^4$Petersburg Nuclear Physics Institute, Gatchina 188300, Russia}


\date{\today}

\begin{abstract}

We optically excite the electronic state $3s3p~^3P_{0}$ in $^{24}$Mg atoms, laser-cooled and trapped in a magic-wavelength lattice. An applied magnetic field enhances the coupling of the light to the otherwise strictly forbidden transition. We determine the magic wavelength, the quadratic magnetic Zeeman shift and the transition frequency to be 468.463(207)$\,\nano\meter$, $-206.6(2.0)\,\mega\hertz$/$\tesla^2$ and 655 058 646 691(101)$\,\kilo\hertz$, respectively. These are compared with theoretical predictions and results from complementary experiments. We also developed a high-precision relativistic structure model for magnesium, give an improved theoretical value for the blackbody radiation shift and discuss a clock based on bosonic magnesium.

\end{abstract}


\maketitle
The frequencies of optical clocks are currently measured with a fractional accuracy and precision of nearly $10^{-18}$~\cite{Chou2010,Hinkley2013,Bloom2014,Ushijima2015}. A potentially limiting systematic frequency shift of both ion and optical lattice clocks is the AC Stark shift from room-temperature black body radiation (BBR)~\cite{NicCamHut15,BelHinPhil14}. Clock transitions with small BBR sensitivities are an attractive approach to even higher accuracies. Among these are neutral Hg and Mg, In$^+$, and especially the Al$^+$ ion clock transition, which all have significantly smaller BBR sensitivities than Sr and Yb lattice clocks and Cs microwave clocks. 

In this Letter we report the spectroscopy of the Mg clock transition in a magic wavelength optical lattice, which gives equal AC Stark shifts of the clock states. We measure the transition frequency~\cite{gordone::magnesium.frequency.standard,Friebe.2011} and its magic wavelength and quadratic Zeeman shift, which were recently predicted~\cite{santra::magic.wavelength,ovsiannikov::magic.wavelength,derevianko::magic.wavelength,taichanechev::magnetic.induced.3P0.mixing}. 

Along with our measurements, we developed a more refined atomic structure model to calculate both the magic wavelength as well as the static BBR shift. For less massive atoms, such as Mg, these models are more accurate than for heavier elements like Sr and Yb, and spectroscopy of low-mass elements generally represents an interesting testbed for validating improved theoretical models~\cite{MitSafCla10}. Both our theoretical and experimental results for the magic wavelength agree at a level of better than 1\% and restrict the value, which was estimated to fall in the range between 466 and 480$\,\nano\meter$~\cite{santra::magic.wavelength,ovsiannikov::magic.wavelength,derevianko::magic.wavelength}. Our model for Mg predicts a static BBR shift to be eight and five times lower than those of Sr and Yb, which were recently measured~\cite{MidFalLis12,SheLemHin12}. Apart from the static contribution, the total BBR shift also includes a dynamic contribution, which is derived from the combination of theoretical calculations and measurements of $^3D_1$ state lifetime \cite{NicCamHut15,BelSheLem12}. Ref.~\cite{Porsev2006} estimated the dynamic contribution in Mg to be 0.1 a.u. for the $^3P_0$ state, being remarkably smaller than for Yb (1.86 a.u.) and Sr (12.37 a.u.).

For bosonic atoms, optical dipole excitation of the electronic ground state $^1S_{0}$ to $^3P_{0}$ is strongly suppressed. A magnetic field enhances the dipole coupling, enabling nHz linewidth by mixing the $^3P_{1}$ electronic state~\cite{taichanechev::magnetic.induced.3P0.mixing,Barber2006}. Refs.~\cite{taichanechev::magnetic.induced.3P0.mixing,Taichenachev} calculated the associated second order Zeeman effect for Mg to be $-217(11)\,\mega\hertz/\tesla^2$ (equivalent to a fractional frequency shift of $-3.31(17)\times 10^{-7}/\tesla^2$), a systematic effect that must be evaluated. We precisely measured the magnetic field dependence, which is consistent within the uncertainty of~\cite{taichanechev::magnetic.induced.3P0.mixing}, estimated to be 5$\,\%$~\cite{Taichenachev}.  This second-order Zeeman shift is larger than those of Yb ($-6.6(4)\,\mega\hertz/\tesla^2$ or $-1.27(8)\times 10^{-8}/\tesla^2$~\cite{Barber2006}) and Sr ($-23.5(2)\,\mega\hertz/\tesla^2$ or $-5.47(47)\times 10^{-8}/\tesla^2$~\cite{Westergaard2011}).

In this way we directly measure the transition frequency, which agrees with the difference of the measured frequencies of the $^1S_{0}-$$^3P_{1}$ and $^3P_{0}-$$^3P_{1}$ transitions~\cite{Friebe.2011,gordone::magnesium.frequency.standard}. Due to its low mass and the short magic wavelength, Mg has a large photon recoil frequency shift, $\Delta\nu_{\textnormal{R}}=h/(\lambda_{\textrm{magic}}m_{{\textnormal{Mg}}})$, where $\lambda_{\textrm{magic}}$ is the magic wavelength, as well as greater tunneling. A deeper lattice is therefore required to suppress tunneling~\cite{Lemonde2005}, as compared to heavier species.

\begin{figure}
                    \includegraphics[width=1\linewidth]{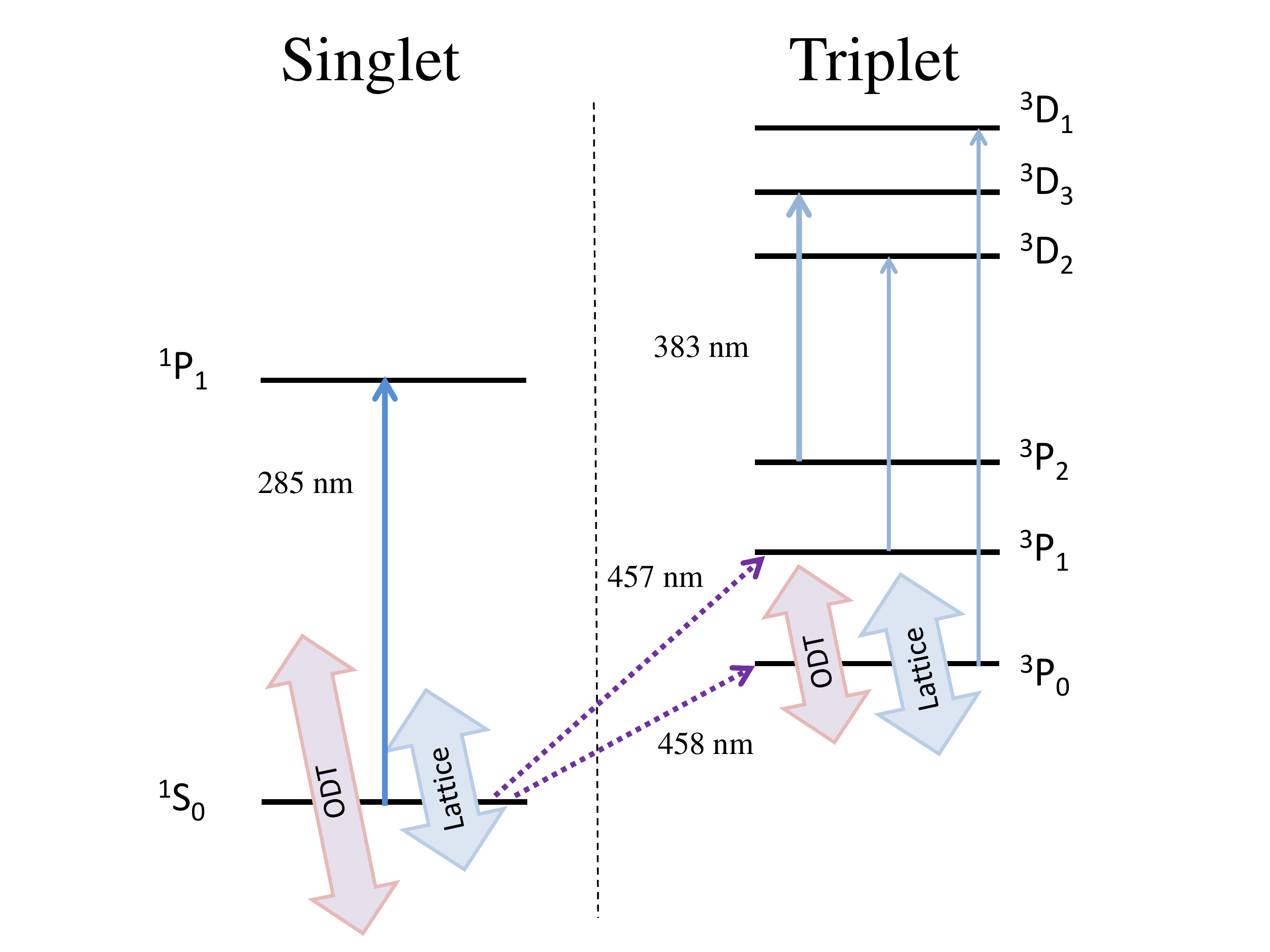}
                    \caption{Optical transitions in $^{24}$Mg relevant for performing optical lattice spectroscopy. Atoms are continuously loaded in the long lived electronic state $^3P_{0}$ in an optical dipole trap (ODT) at 1064$\,\nano\meter$ using a dual magneto-optical trap (MOT)~\cite{Riedmann2012}. Atoms trapped in a MOT using the $^1S_{0}-$$^1P_{1}$ transition are optically transferred with 457 nm to the $^3P_{1}$ state and then to $^3P_{2}$. The atoms are further cooled in a MOT with 383 nm light that excites the $^3D$-manifold, and cold atoms are permitted to accumulate in $^3P_{0}$ in the ODT. These atoms are optically depumped to the ground state via $^3P_{1}$ and the magic wavelength optical lattice is adiabatically turned on. The dipole trap and the optical lattice laser beams are depicted by the bold arrows. The 458$\,\nano\meter$ light interrogates the magnetic field enhanced clock transition.  }
                    \label{levelscheme}
\end{figure}

We briefly summarize the steps required for the optical lattice spectroscopy in Fig. 1. A thermal beam of Mg atoms is slowed and loaded into a "singlet"-magneto-optical trap (MOT) using laser light tuned near the $^1S_{0}-$$^1P_{1}$ transition. Atoms are optically transferred, after excitation to the $^3P_{1}$ state, into a second, simultaneously operated "triplet" $^3P_{2}-$$^3D_3$ MOT. There atoms can decay to the $^3P_{1}$ state (see Fig. \ref{levelscheme}) during MOT operation~\cite{Hansen2008} and have to be recycled with light exciting them to the $^3D_{2}$ state. This yields 10$^5$ atoms in the $^3P_{0}$ state at 100$\,\micro\kelvin$ in a 1064$\,\nano\meter$ dipole trap as in~\cite{Riedmann2012}. The atoms are then optically pumped to the $^3D_{2}$ state and decay to the electronic ground state via the $^3P_{1}$ state. After this, a spatially-overlapped 1D optical lattice with a waist of 89$\,\micro\meter$ is adiabatically turned on before the dipole trap is rapidly switched off. To select the coldest atoms, the optical lattice intensity is ramped down to a certain depth and subsequently increased to a final value for the clock transition spectroscopy. This procedure reproducibly generates about 1000 atoms at 4$\,\micro\kelvin$.

We generate 160$\,\milli\watt$ of lattice light near the magic wavelength at $\lambda_{\textrm{magic}}=469\,\nano\meter$ with a frequency-doubled Ti:Sa laser. A horizontal build-up cavity, with a finesse of 80, twines around the vacuum chamber and, with a circulating power of $\sim$2.3$\,\watt$, produces trap depths of 10 recoil energies $h\nu_{\textnormal{R}}$. The cavity length is stabilized to the frequency of the laser with a Pound-Drever-Hall~\cite{drever.hall::pdh} lock driving an electro-optical modulator (EOM) and a piezo-controlled mirror. An additional feedback loop driving an acousto-optical modulator (AOM) can set and stabilize the depth of the lattice. The light transmitted through the cavity is used to measure the circulating light intensity in the cavity.

The clock transition spectroscopy is performed with a home-built external cavity diode laser stabilized to an ultrastable resonator with $\mathcal{F}=600.000$ at 916$\,\nano\meter$, similar to~\cite{Pape2010}. The infra-red light is fiber-guided to the spectroscopy setup, a tapered amplifier chip, and a commercial second-harmonic generation (SHG) stage. The system generates 10$\,\milli\watt$ of 458$\,\nano\meter$ light with a short-term frequency instability as low as $5\times 10^{-16}$ in 1$\,\second$. The spectroscopy is performed by irradiating the atoms for 100$\,\milli\second$ with a pulsed, Gaussian shaped laser beam with a waist of 300$\,\micro\meter$ and a peak intensity of 7.07$\,\watt/\centi\meter^2$. The MOT coils, operated in Helmholtz configuration, generate a magnetic field of 2.49(1)$\,$G/A, determined via optical Zeeman spectroscopy of the $^1S_{0}(m_J=0)-^{3}P_{1}(m_J=\pm 1)$ transitions, increasing the dipole coupling of $^1S_{0}$ and $^3P_{0}$. We normally use a magnetic field of 249$\,$G, which yields a predicted linewidth of 8.07$\,\micro\hertz$ and a Rabi frequency of 205$\,\hertz$ ~\cite{taichanechev::magnetic.induced.3P0.mixing}. In this way, we resonantly excite up to a thousand atoms to $^3P_{0}$, which are then optically pumped to $^{3}P_{2}$ and detected with a few ms of fluorescence from the "triplet"-MOT. This detection scheme yields a sensitivity of a few tens of atoms. To obtain the line centre and profile of the transition, we record the number of excited atoms as we step the frequency of the 458$\,\nano\meter$ laser.  Using the atoms, the linear drift of the laser is determined at the beginning and end of the spectroscopic measurements. The initial drift is compensated with a feed-forward of an AOM that shifts the laser frequency to a resonance of our ultrastable cavity. A scan over the resonance typically comprises 30 measurements, each lasting 1.9$\,\second$.

\begin{figure*}[t]
                    \includegraphics[width=\textwidth]{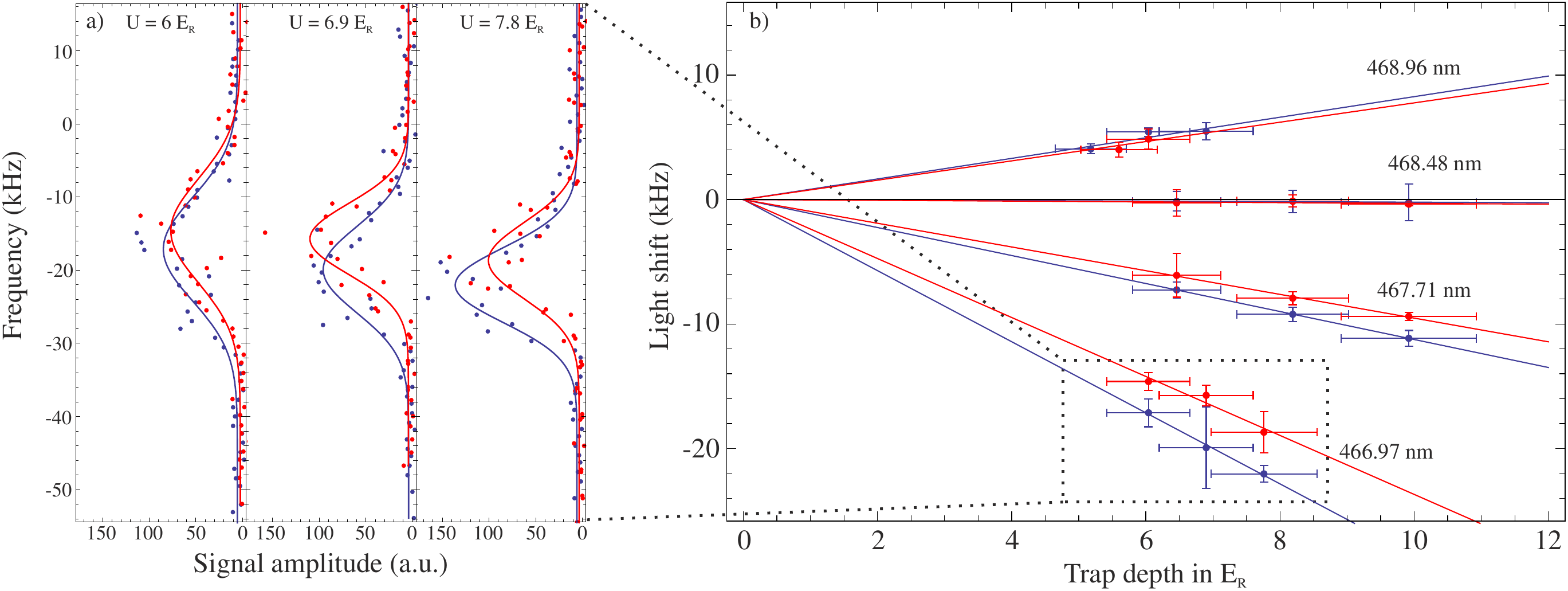}
                    \caption{a) Measured and fit line profiles for three depths of an 466.97$\,\nano\meter$ optical lattice. Two sets of measurements (red and blue dots) are shown with their corresponding Gaussian fits (red and blue solid lines). b) The observed linear AC Stark shift versus optical lattice depth for several lattice wavelengths. The frequencies of the line center (dots) from Gaussian fits as in a) and their corresponding linear regression (solid lines) are depicted. For each measurement set at a lattice wavelength, a single frequency offset accounts for the drift of our ultrastable cavity.}
                    \label{lightshift}
\end{figure*}

The magic wavelength for $^{24}$Mg is inferred from measurements of the line centre for different lattice depths and several wavelengths. Figure \ref{lightshift} a) shows two sets of measurements of the transition probability (red and blue dots) versus clock laser frequency and corresponding Gaussian fits (red and blue solid curves) for three depths of a 466.97$\,\nano\meter$ lattice. The line profiles for different trap depths were measured successively. To evaluate and correct the residual laser drift, the measurement sequence was repeated three times and the shift of the line centers for a specific trap depth is determined from the Gaussian fits. From the frequencies for a specific lattice depth, we infer the residual clock laser drift, which can be as large as 2-3$\,\kilo\hertz$ within several minutes. The line profiles in Figure \ref{lightshift} a) are three superposed scans. The linewidth of each profile, on the order of a few kHz, is mostly due to tunneling in our shallow optical lattice. Figure \ref{lightshift} b) shows the  line centers (dots) and the corresponding linear regression (solid lines) of the AC Stark shifts versus lattice depth for different lattice wavelengths. An offset frequency was subtracted from the linear regressions for each lattice wavelength to account for the laser drift between measurements. The uncertainty of the experimental determination of the lattice depth is about 5$\,\%$ dominated by the uncertainty of the lattice waist. The uncertainty of the AC Stark shift is a combination of the statistical uncertainty of the linear regression and the systematic uncertainty of residual (non-linear) frequency drifts of the clock laser, on the order of a few kHz. The differences in the linear regression from the two measurement campaigns agree within these uncertainties. Separately, the two data sets yield magic wavelengths of 468.472(224)$\,\nano\meter$ (blue data) and 468.452(192)$\,\nano\meter$ (red data). Applying a linear regression to the combination of both measurement sets, we determine the magic wavelength of the $^{24}$Mg $^1S_{0}-$$^3P_{0}$ transition to be 468.463(207)$\,\nano\meter$ and the linear AC Stark shift dependence on lattice depth and wavelength to be $1.669(115)\,\kilo\hertz/\textnormal{E}_{\textnormal{R}}/\nano\meter$ (equivalent to a fractional frequency shift of $2.55(18)\times 10^{-12}/\textnormal{E}_{\textnormal{R}}/\nano\meter$).

  %MSS
The experimental results agree well with our theoretical model using a state-of-the-art relativistic approach that combines configuration interaction and all-order linearized coupled-cluster methods (CI+all-order). Our final recommended value for the theoretical Mg magic wavelength includes the replacement of the calculated values of the transition energies of the dominant contributions by experimental values. While our calculated Mg transition energies agree with the observed values to a few cm$^{-1}$, even these small differences affect the magic wavelength in the fourth significant figure.  To evaluate the uncertainty of our theoretical calculations, we carried out the calculations using a combination of the CI and second-order many body perturbations theory (CI+MBPT), which does not include all-order corrections to the effective Hamiltonian. The difference of the CI+MBPT and CI+all-order values serves as an estimate of the theoretical accuracy ~\cite{Safronova2011,Safronova2012,Safronova2013}.
The results are summarized in Table~\ref{tab1}. We give CI+MBPT and CI+all-order values for the magic wavelength $\lambda_{\textrm{magic}}$, as well as the static ground state $\alpha(ns^2$~$^1S_0)$ and excited clock state $\alpha(nsnp~^3P_0)$ polarizabilities, and their difference $\Delta \alpha$, which is proportional to the static BBR shift~\cite{Porsev2006}. To demonstrate the extremely high accuracy of the theoretical calculations in Mg, we compare the magic wavelength and polarizabilities of Mg, Sr, and  Yb in Table~\ref{tab1}. The large differences between CI+MBPT and CI+all-order Sr and Yb values illustrate the significance of higher-order effects in these heavier systems. The excellent agreement of the CI+MBPT and CI+all-order polarizabilities indicates an uncertainty of the Mg BBR shift of less than 1\%.
   % High accuracy of the theoretical calculations will aid further development of a Mg lattice clock.
 \begin{table}
\caption{\label{tab1}
Comparison of CI+MBPT and CI+all-order values for magic wavelengths $\lambda_{\textrm{magic}}$ in nm  and static polarizabilities $\alpha$ (in a.u.) of Mg, Sr \cite{Safronova2013}, and Yb \cite{Safronova2012}.
$\Delta \alpha=\alpha(nsnp~^3P_0)-\alpha(ns^2$~$^1S_0)$. $^a$We only list 6 significant figures from the measurements in \cite{LudZelCam08,BarStaLem08}.
$^b$Using experimental energies gives 254.4~a.u. and small corrections yield 247.5~a.u.}
\begin{ruledtabular}
\begin{tabular}{llccc}
\multicolumn{1}{c}{Quantity}&
\multicolumn{1}{c}{Method}&
\multicolumn{1}{c}{Mg}&
\multicolumn{1}{c}{Sr}&
\multicolumn{1}{c}{Yb}\\
\hline
$\lambda_{\textrm{magic}}$                &CI+MBPT          &   468.45        &   847          &  789     \\
                                          &CI+all     &   468.68        &   820       &   754    \\
                                          &Final&   468.45(23)    &             &         \\
                                          &Expt.            &   468.463(207)  &   813.427$^a$     &  759.354$^a$ \\
$\alpha(^1S_0)$     &CI+MBPT          &   71.257        &   195.4     &  138.3  \\
                                          &CI+all     &   71.251        &   197.8     &  140.9  \\
$\alpha(^3P_0)$       &CI+MBPT          &   100.812       &   482.1     &  305.9 \\
                                          &CI+all     &   100.922       &   458.1     &  293.2 \\
$\Delta \alpha$ &CI+all     &   29.671        &   260.3$^b$    &  152.3 \\
                                          &Expt.            &                 &   247.379(7)&  145.726(3)\\
    \end{tabular}
\end{ruledtabular}
\end{table}
%MSS******= Sr 813.42735(40)nm Yb: 759.3537(40)nm

\begin{figure}[t]
                    \includegraphics[width=\linewidth]{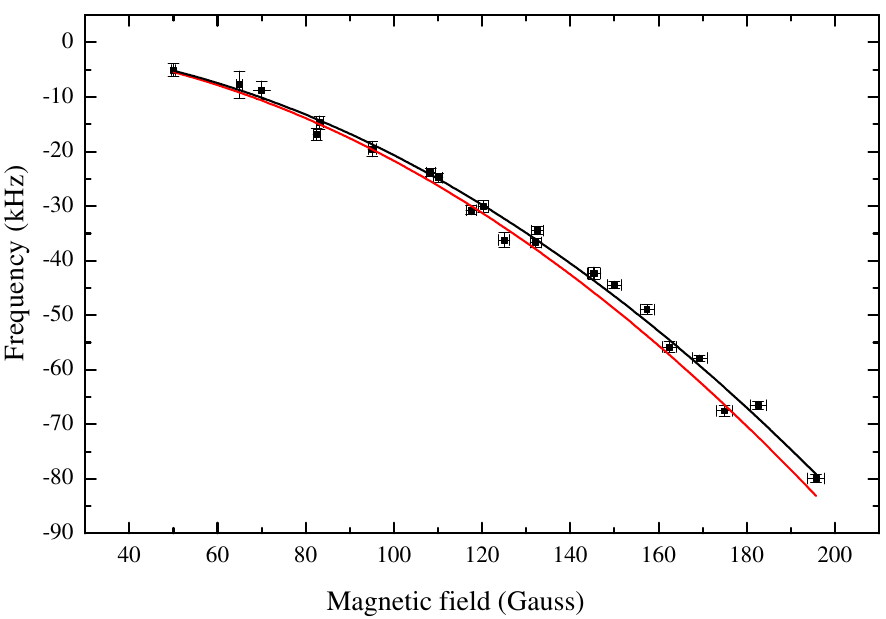}
                    \caption{Quadratic Zeeman shift of the clock transition versus magnetic field strength (black squares), a parabolic fit (black curve), and the theoretical prediction of \cite{taichanechev::magnetic.induced.3P0.mixing} (red curve). The predicted dependence on the magnetic field is -217(11)$\,\mega\hertz/\tesla^2$, which agrees with the experimental result of -206.6(2.0)$\,\mega\hertz/\tesla^2$.} 
                    \label{zeeman}
\end{figure}

The second order Zeeman shift drops out of the determination of the magic wavelength but is a significant correction to our measured transition frequency. Fig. \ref{zeeman} shows the measured transition frequency versus applied magnetic field (squares), a parabolic fit (black curve) of the measurements, and a theoretical prediction by \cite{taichanechev::magnetic.induced.3P0.mixing} (red curve). The experimental parabolic coefficient is \mbox{$-206.6(2.0)\,\mega\hertz/\tesla^2$} (equivalent to $-3.15(3)\times 10^{-7}/\tesla^2$) and agrees within 5$\,\%$ with a theoretical value, which is consistent with its estimated uncertainty ~\cite{Taichenachev}. The measurement accuracy of the magnetic field strength, via the Zeeman spectroscopy of the $^1S_{0}-$$^3P_{0}$ transition, is 1$\,\%$ and limited by our present accuracy in measuring the MOT coils' electrical current.

Our measurements at the magic wavelength, with the correction of the second-order Zeeman shift, yield a direct measurement of the optical transition frequency of 655 058 646 691(101)$\,\kilo\hertz$. The absolute frequency is measured by beating the spectroscopy laser with an optical frequency comb that is stabilized to a 10$\,\mega\hertz$ GPS frequency reference. Within the dominant uncertainty from the GPS reference, the transition frequency agrees with the difference of previous spectroscopic measurements of the $^1S_{0}-$$^3P_{1}$ and $^3P_{0}-$$^3P_{1}$ transitions of 655 659 923 839 730(48)$\hertz$ and 601 277 157 870.0(0.1)$\,\hertz$~\cite{Friebe.2011,gordone::magnesium.frequency.standard}. 

In summary, we report the direct optical spectroscopy of the $^1S_{0}-$$^3P_{0}$ clock transition of laser cooled bosonic $^{24}$Mg in a magic-wavelength optical lattice. Our measurements determine precisely the magic wavelength and confirm the high precision obtained with a new theoretical atomic model of Mg. Our experimental determination of the quadratic Zeeman effect and clock transition frequency agree with a prediction~\cite{taichanechev::magnetic.induced.3P0.mixing} and previous indirect frequency measurements. Planned future spectroscopy in a deeper lattice of more than 40 recoil energies, will reduce the width of the lowest vibrational band to $\sim\,$20$\,\hertz$ and thus allow high clock accuracies.  

The demonstrated agreement of our combination of theory and experimental measurements is an important ingredient for exploring a future bosonic and fermionic Mg optical lattice clock. For bosonic magnesium, atoms can be optically prepared at $\micro\kelvin$ temperatures, which has not yet been demonstrated for the fermionic isotope $^{25}$Mg. In our experiment, a dilute atomic cloud of 1000 atoms is distributed over 130.000 lattice sites ($\sim 0.008$ atoms per lattice site), which is a factor of 100 lower density than that reported for other clocks with approximately the same number of atoms~\cite{Ludlow2011,Falke2014}, significantly reducing the limitations from collisional shifts. The second order Zeeman shift can be sufficiently controlled~\cite{footnote1}, especially with higher clock laser intensity. A clock laser intensity of 7.07$\,\watt/\centi\meter^2$ will yield a reasonable Rabi frequency of 20.5$\,\hertz$ requiring a 10 times smaller magnetic field, with a corresponding reduction in the uncertainty of the quadratic Zeeman shift. Further, magnesium offers suitable narrow transitions for precise Zeeman spectroscopy to calibrate the magnetic field. These techniques can exploit the small blackbody radiation shift to make highly accurate and stable lattice clocks and further precisely test atomic models for precision spectroscopy. 
  
We thank A. Bauch and H. Schnatz from Physikalisch Technische Bundesanstalt (PTB) for providing a passive H-maser and assistance for the frequency measurement during the early stage of the experiment as well as C. Lisdat for helpful discussions.

This work was supported in part by Deutsche Forschungsgemeinschaft within QUEST, Center for Quantum Engineering and Space-Time Research (APK, SR, SS, WE and EMR), NASA (KG) and U. S. NSF Grant Nos. PHY-1404156 (MSS and SGP) and PHY-1311570 (KG). DF and KHZ acknowledge financial support from Research Training Group RTG1729. NJ is supported by the Marie Curie Initial Training Networks (ITN) Program, call: FP7-PEOPLE-2013-ITN.

\bibliography{bibliography}

\end{document}